\documentclass[aps,prb,twocolumn,floatfix,showpacs,citeautoscript,longbibliography]{revtex4-1}

\usepackage{amsmath}
\usepackage{amssymb}
\usepackage{amsfonts}
\usepackage{graphicx,graphics}
\usepackage{color}

\newcommand{\ket}[1]{\lvert #1\rangle}
\newcommand{\bra}[1]{\langle#1 \rvert}
\newcommand{\abs}[1]{\lvert #1 \rvert}

\definecolor{ThesisBlue}{RGB}{40,140,170}

\begin{document}

\title{Thermoelectrics in Coulomb-coupled quantum dots:\\ Cotunneling and energy-dependent lead couplings}
\author{Nicklas Walldorf}
\email{nicwall@nanotech.dtu.dk}
\author{Antti-Pekka Jauho}
\author{Kristen Kaasbjerg}
\email{kkaa@nanotech.dtu.dk}
\affiliation{Center for Nanostructured Graphene (CNG), Dept. of Micro- and
  Nanotechnology, Technical University of Denmark, DK-2800 Kongens
  Lyngby, Denmark}
\date{\today}

\begin{abstract}
  We study thermoelectric effects in Coulomb-coupled quantum-dot (CCQD)
  systems beyond lowest-order tunneling processes and the often applied
  wide-band approximation. To this end, we present a master-equation (ME)
  approach based on a perturbative $T$-matrix calculation of the charge
    and heat tunneling rates and transport currents. Applying the method to
  transport through a non-interacting single-level QD, we demonstrate excellent
  agreement with the Landauer-B{\"u}ttiker theory when higher-order
    (cotunneling) processes are included in the ME. Next, we study the effect of
  cotunneling and energy-dependent lead couplings on the heat currents in a
  system of two Coulomb-coupled QDs. Overall, we find that cotunneling processes
  (i) dominate the heat currents at low temperature and bias, and (ii) give rise
  to a pronounced reduction of the cooling power achievable with the
  recently demonstrated Maxwell's demon cooling mechanism. Furthermore, we
  demonstrate that the cooling power can be boosted significantly by carefully engineering the energy dependence of the lead couplings to filter out undesired transport processes. Our findings emphasize
  the importance of considering higher-order cotunneling processes as well as the advantage of
  engineered energy-dependent lead couplings in the optimization of the
  thermoelectric performance of Coulomb-coupled QD systems.
\end{abstract}
\maketitle

\section{Introduction}
\label{sec:Intro}

The experimental progress in control of single-electron
transport~\cite{Pekola:Single} has spurred interest in nanosystems which
utilize the associated heat currents for thermoelectric
applications~\cite{Pekola:Opportunities,Dresselhaus:Thermoelectric,Molenkamp:Thermo}.
In particular, experiments with Coulomb-coupled quantum-dot (CCQD) systems have demonstrated a plethora of novel phenomena
ranging from Coulomb drag~\cite{Ensslin:Measurement,Gordon:Cotunneling} and electron
pairing~\cite{Ilani:Attraction} to extraordinary thermoelectric
effects~\cite{Molenkamp:Three,Pekola:Onchip}. This includes the realization
of an energy harvester which converts a thermal gradient in a CCQD system into an electric current~\cite{Molenkamp:Three}, as well as an autonomous
Maxwell's demon capable of cooling a current-carrying QD system at the cost of
heating a ``demon'' QD system~\cite{Pekola:Onchip}.

\begin{figure}[!b]
  \centering
  \includegraphics[width=0.85\columnwidth]{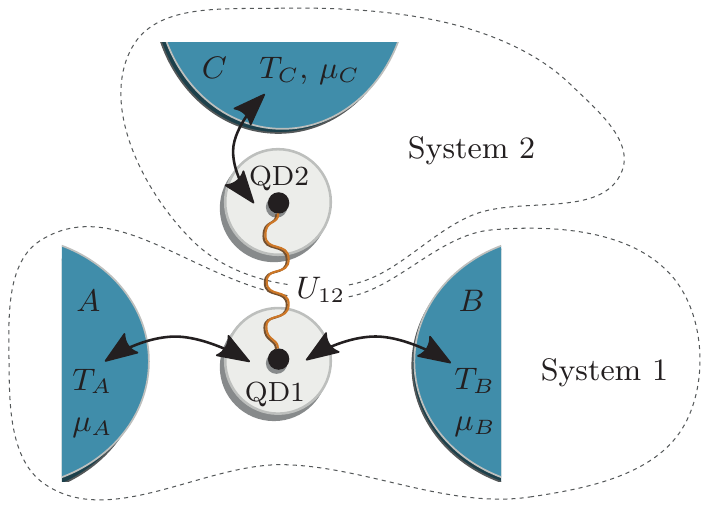}
  \caption{Illustration of the CCQD system studied in Sec.~\ref{sec:CQD}
      consisting of two Coulomb-coupled QDs
      $\delta\in\{1,2\}$ with inter-dot Coulomb interaction
      $U_{12}$, tunnel-coupled in a three-terminal configuration to leads
    $\ell\in\{A,B,C\}$ (no tunneling allowed between the QDs) with temperatures $T_\ell$ and electrochemical potentials
    $\mu_\ell$.}
\label{fig:System}
\end{figure}

In addition to the above, theoretical studies have predicted a wide range of
novel thermoelectric effects in CCQD
systems~\cite{Strasberg:Thermodynamics,
Buttiker:Optimal,Jordan:Thermoelectric,Sanchez:All}.
The mechanisms behind these effects rely on the presence of a
strong Coulomb interaction between electrons in the otherwise decoupled QDs
(see Fig.~\ref{fig:System} for the case of two Coulomb-coupled QDs). The strong interaction can be utilized to tailor the thermoelectric properties of CCQD
systems~\cite{Molenkamp:Thermo,Koski:Maxwell} and opens the opportunity to test
fundamental thermodynamic aspects of heat transport in interacting
nanoscale systems driven out of
equilibrium~\cite{Whitney:Fundamental}.

While the operation principles of the above-mentioned effects are
governed by \emph{incoherent} electron tunneling (sequential
tunneling) processes between the leads and the
QDs~\cite{Strasberg:Thermodynamics,Buttiker:Optimal,Jordan:Thermoelectric,Molenkamp:Three,Pekola:Onchip,Sanchez:All},
the importance of coherent higher-order tunneling (cotunneling) processes
for the nonlinear heat transport remains largely
unexplored~\cite{Pekola:Onchip}. Furthermore, when operated under strong non-equilibrium conditions where linear response theory
breaks down, a theoretical treatment taking into account
the full nonlinear
properties is needed~\cite{Haupt:Heat,Leijnse:Nonlinear,Lopez:Heat,Lopez:Nonlinear}. Only
recently have these issues been discussed in strongly interacting QD
systems~\cite{Schuricht:Charge,Seja:Violation,Dare:Powerful}.

Another important factor for thermoelectric effects in CCQD systems is the coupling to the leads which is usually treated in
the wide-band approximation assuming energy-independent
couplings~\cite{Flensberg}. However, energy-dependent couplings to the leads
occur naturally in many QD 
systems~\cite{Ilani:Realization,Molenkamp:Three,Ensslin:Measurement,Gordon:Cotunneling}, and add an
important degree of tunability to the system, and is as crucial for the
thermoelectric properties~\cite{Strasberg:Thermodynamics,Buttiker:Optimal,Zhang:Three} as it is for
Coulomb
drag~\cite{Buttiker:Mesoscopic,Ensslin:Measurement,Jauho:Correlated,Gordon:Cotunneling,Lim:Engineering}.

In this work, we present a master-equation approach for the calculation of the
nonlinear electronic charge and heat currents in interacting QD systems
which takes into account the above-mentioned factors. The charge and heat
transfer rates produced by electron tunneling processes are obtained with a
perturbative $T$-matrix approach~\cite{Flensberg} which allows us to treat
sequential and cotunneling processes on equal footing. We resolve the technical
challenges associated with the evaluation of the cotunneling rates with an
implementation of the often applied regularization
scheme~\cite{Matveev:Cotunneling,Koch:Cotunneling} which applies to the general
case of energy-dependent lead couplings, applied biases, and
temperature gradients in the system.

The main findings and the organization of the paper are as follows. In
Sec. \ref{sec:Model}, we introduce the model system of CCQDs. In
Sec. \ref{sec:Master}, we present the methodology, and benchmark the approach in
Sec. \ref{sec:SingleLevel} by comparing it to the Landauer-B{\"u}ttiker
formalism for transport through a non-interacting single-level QD. In
Sec. \ref{sec:CQD}, we study nonlinear thermoelectric phenomena in CCQDs. We
investigate the energy exchange mediated by the inter-dot Coulomb interaction
which among other thermoelectric effects leads to the demon-induced cooling
mechanism~\cite{Strasberg:Thermodynamics,Pekola:Onchip}. Our findings shed light on the limitations imposed by cotunneling processes on the
performance of this mechanism. Furthermore, we demonstrate a strongly enhanced
performance of the demon-induced cooling effect by tuning the
energy-dependence of the lead couplings. In such performance optimization, as
we show, cotunneling processes are essential for a quantitative description of
the thermoelectric properties. Finally, Sec. \ref{sec:conc} presents our
conclusions, and App. \ref{sec:regularization} gives technical details on the
cotunneling rates and the regularization procedure.

\section{Coulomb-coupled QD systems}
\label{sec:Model}

We consider CCQD systems like the one
illustrated in Fig.~\ref{fig:System}, which can be described by the Hamiltonian
\begin{equation}
  \hat{H} = \hat{H}_\text{dots} + \hat{H}_\text{leads} + \hat{H}_T,
\end{equation}
and consists of a system of CCQDs with
Hamiltonian $\hat{H}_\text{dots}$ which is coupled to external leads with Hamiltonian
$\hat{H}_\text{leads}$ by tunnel couplings described by $\hat{H}_T$. We denote
$\hat{H}_0=\hat{H}_\text{dots} + \hat{H}_\text{leads}$.

For the QD system, we consider a minimal spinless model of inter-dot
Coulomb-coupled single-level QDs described by the Hamiltonian
\begin{equation}
  \label{eq:Hdots}
\hat{H}_\text{dots}=\sum_{\delta}\epsilon_\delta^{\phantom\dagger}\hat{c}_\delta^\dagger \hat{c}_\delta^{\phantom\dagger}+\sum_{\langle\delta,\delta'\rangle}U^{\phantom\dagger}_{\delta\delta'}\hat{n}^{\phantom\dagger}_{\delta}\hat{n}^{\phantom\dagger}_{\delta'},
\end{equation}
where $\hat{c}_\delta^\dagger$ ($\hat{c}_\delta^{\phantom\dagger}$) creates (annihilates) an
electron in QD $\delta$ with energy controlled by gate voltages $\epsilon_\delta=-eV_\delta$, where $V_\delta$ is
the gate potential on dot $\delta$,
$\hat{n}^{\phantom\dagger}_\delta=\hat{c}_{\delta}^\dagger \hat{c}^{\phantom\dagger}_{\delta}$ is
the occupation number operator, $U_{\delta\delta'}$ is the inter-dot Coulomb
interaction, and the summation in the second term is over all QD pairs (specific systems are studied in Secs. \ref{sec:SingleLevel}--\ref{sec:CQD}).

The leads are described by non-interacting electron reservoirs,
  $\hat{H}_\text{leads}=\sum_{\ell k}\epsilon_{\ell k}^{\phantom\dagger}\hat{c}_{\ell k}^\dagger \hat{c}_{\ell k}^{\phantom\dagger}$,
where $\hat{c}_{\ell k}^\dagger$ ($\hat{c}_{\ell k}^{\phantom\dagger}$) creates
(annihilates) an electron with momentum $k$ and energy $\epsilon_{\ell k}$ in lead
$\ell$, which is assumed to be in local equilibrium with temperature $T_\ell$ and
electrochemical potential $\mu_\ell=\mu_0-eV_\ell$, where $\mu_0$ is the
equilibrium chemical potential and $V_\ell$ is the voltage applied to lead $\ell$. The tunneling Hamiltonian which couples the QD system to the leads is 
$\hat{H}_T=\sum_{\ell k \delta}(t^{\phantom\dagger}_{\ell k\delta}\hat{c}^\dagger_\delta \hat{c}^{\phantom\dagger}_{\ell k}+\textit{h.c.}),$
where $t_{\ell k\delta}$ is the tunneling amplitude. We define lead coupling
strengths as
$\gamma^\ell(\epsilon) \equiv 2\pi d_\ell(\epsilon)|t_\ell(\epsilon)|^2$, where
$d_\ell(\epsilon)$ is the lead density of states. $\gamma^\ell(\epsilon)$ is allowed to be energy dependent
in contrast to the often applied wide-band approximation.

\section{Master equation and transport currents}
\label{sec:Master}

We describe the dynamics and transport in the CCQD system with a Pauli ME where the
transitions between the QD states are governed by electron tunneling to and from
the leads~\cite{Timm:Tunneling}. The tunneling-induced transition rates are
calculated based on a perturbative $T$-matrix approach where the tunneling Hamiltonian is
treated as a perturbation to the decoupled QD system and leads. This allows a systematic expansion in the tunnel couplings and the inclusion of high-order
processes. However, quantum effects such as tunneling-induced level broadening
and level shifts~\cite{Konig:Resonant,Schon:Cotunneling,Jonas:Tunneling} are not captured by
this perturbative approach, which is only valid in the weak coupling regime
$\gamma< k_BT, U$.

In the absence of tunnel coupling, the states of the decoupled QD system and
leads are described by product states of the QD system occupation states $\ket{m}$ with energy
$E_{\text{dots},m}=\bra{m}\hat{H}_\text{dots}\ket{m}$ and the leads $\ket{i}$ with
energy $E_{\text{leads},i}=\bra{i}\hat{H}_\text{leads}\ket{i}$. The non-equilibrium
occupations of the QD states are described by probabilities $p_m$ (the diagonal components of the reduced density operator of the CCQD system) which are
determined by the ME
\begin{equation}\label{eq:master}
  \dot{p}_m = \sum_{n\neq m}\left(\Gamma_{nm}p_n-\Gamma_{mn}p_m\right),\quad   \sum_{m}p_m=1,
\end{equation}
where $\Gamma_{mn}$ denotes the tunneling-induced transition rate from QD state
$\ket{m}$ to $\ket{n}$. The ME is solved for the steady-state
probabilities, $\dot{p}_m=0$, in the following. The QD states are given explicitly
in Sec. \ref{sec:SingleLevel} and Sec. \ref{sec:CQD} for the considered systems.

\subsection{Transition rates}

The rates for transitions between the QD states are obtained from the
generalized Fermi's golden rule~\cite{Flensberg,Timm:Time}
\begin{equation}\label{GFGR}
\tilde{\Gamma}_{mn} = \frac{2\pi}{\hbar}\!\sum_{ij}\!|\bra{j}\bra{n}T\ket{m}\ket{i}|^2\rho_i\delta(\Delta_{mn}+E_{\text{leads},j}-E_{\text{leads},i}),
\end{equation}
where $\Delta_{mn}\equiv E_{\text{dots},n}-E_{\text{dots},m}$, $\rho_i$ is the
thermal probability of finding the leads in the initial state, the sum is over
initial and final states of the leads, and the $T$ matrix obeys
\begin{equation}
\hat{T}=\hat{H}_T+\hat{H}_T\frac{1}{E_\text{initial}-\hat{H}_0+i\eta}\hat{T},
\end{equation}
with $E_\text{initial}=E_{\text{dots},m}+E_{\text{leads},i}$, and $\eta$ is a
positive infinitesimal.

The lowest-order contribution to the tunneling rates describes single-electron
tunneling, or \textit{sequential tunneling}, processes between the QD system and
the leads:
\begin{align}
\Gamma_{mn}^{\small{\overrightarrow{\ell}}}&=\hbar^{-1}\gamma^\ell(\Delta_{mn})f^\ell(\Delta_{mn}),\label{SeqTuRates1}\\
\Gamma_{mn}^{\small{\overleftarrow{\ell}}}&=\hbar^{-1}\gamma^\ell(\Delta_{nm})\bar{f}^\ell(\Delta_{nm})\label{SeqTuRates2},
\end{align}
where Eq. \eqref{SeqTuRates1} (Eq. \eqref{SeqTuRates2}) is the sequential rate of tunneling out of,
$\rightarrow$, (into, $\leftarrow$) lead $\ell$, thereby changing the state of the QD system from $m$ to $n$,
${f^\ell(\epsilon)=[\exp{(\beta_\ell(\epsilon-\mu_\ell))}+1]^{-1}}$ is the
Fermi-Dirac distribution in lead $\ell$, $\bar{f}^\ell(\epsilon)=1-f^\ell(\epsilon)$, and
$\beta_\ell=1/(k_BT_\ell)$. The leads are assumed to equilibrate to the Fermi-Dirac distribution in between the tunneling events.

The next-to-leading order terms in the $T$ matrix describe cotunneling
processes. In conventional \textit{local} elastic and inelastic cotunneling
processes, a net electron is transferred between two leads attached to the same
QD (e.g., System 1 in Fig.~\ref{fig:System}). Here we also consider (i)
\emph{nonlocal} cotunneling
processes~\cite{Gordon:PseudoSpin,Jauho:Correlated} in which a net electron is transferred between leads attached to different QDs, as well as (ii)
pair-cotunneling processes where two electrons tunnel into/out of the CCQD
  system in one coherent process~\cite{Koch:Pair,Leijnse:Pair}.

For the thermoelectric effects in focus here, the process of nonlocal
cotunneling is important. The (unregularized) rate for nonlocal cotunneling
which net transfers an electron out of lead $\ell$ and into lead $\ell'$ is given by
\begin{align}
  \label{eq:Cotunneling1}
  \tilde{\Gamma}^{\small{\overrightarrow{\ell\phantom{'}}\overleftarrow{\ell'}}}_{mn}
  & =  \int \!\frac{d\epsilon}{2\pi\hbar}
       \gamma^{\ell}(\epsilon) \gamma^{\ell'}\!(\epsilon - \Delta_{mn})    
       f^{\ell\phantom{'}}\!(\epsilon)\bar{f}^{\ell'}(\epsilon -\Delta_{mn})
    \nonumber\\
  &\quad\times \left|\frac{1}{\Delta_{vm} + \epsilon + i\eta} 
    + \frac{1}{\Delta_{v'n} - \epsilon + i\eta} \right|^2 ,
\end{align}
where $v$ ($v'$) refers to the virtually occupied intermediate state in the process where an
electron initially tunnels from lead $\ell$ and into the QD system (from the QD
system and into lead $\ell'$). We refer to App. \ref{sec:regularization} for the
expressions for the remaining cotunneling processes relevant for this study.

A well-known artifact of the cotunneling rates obtained with the $T$-matrix approach is that they formally diverge in
the limit $\eta\to 0$. To deal with this divergence different regularization schemes have been proposed\cite{Matveev:Cotunneling,Koch:Cotunneling,Koller:Density,Timm:Time}. Deep inside the Coulomb blockade, the discripancy between the different regularization schemes vanishes\cite{Koller:Density}. In this work, we apply the by now standard regularization scheme in Ref.~\onlinecite{Matveev:Cotunneling}, but for future work, a detailed comparison of the charge and heat currents obtained from different regularization schemes could be useful. We
denote the regularized rates which enter into Eq. \eqref{eq:master} without a tilde. To be explicit, we consider the processes
$
\Gamma^{\phantom{\ell\rightarrow}}_{mn}\equiv\sum_{\ell}(\Gamma_{mn}^{\ell\leftarrow}+\Gamma_{mn}^{\ell\rightarrow}),\
\Gamma^{\ell\leftarrow}_{mn}\equiv
\Gamma^{\small{\overleftarrow{\ell\phantom{'}}}}_{mn}+\sum_{\ell'}(\Gamma^{\small{\overleftarrow{\ell\phantom{'}}\overrightarrow{\ell'}}}_{mn}+\Gamma^{\small{\overleftarrow{\ell\phantom{'}}\overleftarrow{\ell'}}}_{mn}),\
\Gamma^{\ell\rightarrow}_{mn}\equiv
\Gamma^{\small{\overrightarrow{\ell\phantom{'}}}}_{mn}+\sum_{\ell'}(\Gamma^{\small{\overrightarrow{\ell\phantom{'}}\overleftarrow{\ell'}}}_{mn}+\Gamma^{\small{\overrightarrow{\ell\phantom{'}}\overrightarrow{\ell'}}}_{mn}).
$ A numerical
procedure for the regularization is outlined in App.~\ref{sec:regularization}.

\subsection{Charge and heat currents}

The steady-state transport currents can be obtained from the occupation
probabilities. The electric current going into lead $\ell$ is
\begin{equation}
\label{eq:Current}
I_\ell\equiv-e\left\langle\sum_k\frac{d\hat{n}_{\ell k}}{dt}\right\rangle=-e\sum_{mn}p^{\phantom{\ell\rightarrow}}_m\!\!\!\left(\Gamma^{\ell\leftarrow}_{mn}-\Gamma^{\ell\rightarrow}_{mn}\right),
\end{equation}
where
$\hat{n}_{\ell k}^{\phantom\dagger}=\hat{c}_{\ell k}^\dagger \hat{c}_{\ell
  k}^{\phantom\dagger}$,
$p_m$ is calculated from Eq. \eqref{eq:master}, and the rightmost form expresses
the electric current in terms of the total rate of electrons tunneling into lead $\ell$, minus the total rate
of electrons tunneling out of lead $\ell$\cite{FN2}.

The heat current going into lead $\ell$
is~\cite{Leijnse:Nonlinear,Sanchez:Dynamics,Whitney:Fundamental}
\begin{equation}
\label{eq:Heat}
J_\ell\!\equiv\!\left\langle\sum_k(\epsilon_{\ell k} - \mu_\ell)\frac{d\hat{n}_{\ell k}}{dt}\right\rangle = \sum_{mn}p^{\phantom{\ell\rightarrow}}_m\!\!\!\left(W^{\ell\leftarrow}_{mn} - W^{\ell\rightarrow}_{mn}\right)\!,
\end{equation}
where the rightmost form expresses the heat current in terms of heat rates $W$
(using a similar notation as for the tunneling rates).

The sequential-tunneling heat rate in lead $\ell$ is calculated as the
tunneling rate multiplied by the energy of the tunneling electron relative to the chemical potential in the lead,
\begin{equation}
\begin{split}
\label{SeqHeRates}
W_{\,\ell,mn}^{\small{\overrightarrow{\ell}}}&=(\Delta_{mn}-\mu_\ell)\Gamma_{mn}^{\small{\overrightarrow{\ell}}},\\
W_{\,\ell,mn}^{\small{\overleftarrow{\ell}}}&=(\Delta_{nm}-\mu_\ell)\Gamma_{mn}^{\small{\overleftarrow{\ell}}},
\end{split}
\end{equation}
where the indices follow the notation of the tunneling rates, however, the
additional first subscript $\ell$ refers to the lead in which the heat rate is
calculated.

Analogously, the cotunneling heat rates into/out of the leads are calculated
  \textit{a posteriori} by multiplying the integrand in the cotunneling rate by
the energy of the tunneling electron relative to the chemical potential of the
lead. For example, for the nonlocal cotunneling process between lead $\ell$ and $\ell'$,
the heat rate in lead $\ell$ reads
\begin{align}
  \label{eq:CotunnelingHeat1}
  \tilde{W}^{\small{\overrightarrow{\ell\phantom{'}}\overleftarrow{\ell'}}}_{\ell,mn}
  & = \!\int\!\!\frac{d\epsilon}{2\pi\hbar}
      \gamma^{\ell}(\epsilon) \gamma^{\ell'}\!(\epsilon - \Delta_{mn}) 
       f^{\ell\phantom{'}}\!(\epsilon) \bar{f}^{\ell'} (\epsilon - \Delta_{mn})
    \nonumber\\
  &\quad\times (\epsilon - \mu_\ell) 
       \left|\frac{1}{\Delta_{vm} +\epsilon + i\eta} +    
             \frac{1}{\Delta_{v'n} - \epsilon + i\eta}
       \right|^2\!\!,
\end{align}
with the heat rate in lead $\ell'$,
$\tilde{W}^{\small{\overrightarrow{\ell\phantom{'}}\overleftarrow{\ell'}}}_{\ell',mn}$,
given as above but with $(\epsilon-\mu_\ell)$ replaced by
$(\epsilon-\Delta_{mn}-\mu_{\ell'})$. The remaining cotunneling heat rates 
follow similarly.

Whereas the calculation of charge currents involves the
electron-tunneling rates which enter the ME~\eqref{eq:master}, and therefore does not require any additional steps once the ME has been set
up and solved, the heat currents must be calculated via the heat tunneling rates in
a post-processing step, similar to more rigorous density-matrix treatments.~\cite{Schuricht:Charge}

\section{Comparison to the Landauer-B{\"u}ttiker formalism}\label{sec:SingleLevel}

In this section, we benchmark the approach by comparing the charge and heat
currents in a spinless non-interacting single-level QD system with those obtained
from the Landauer-B{\"u}ttiker (LB) formalism (see Ref. \onlinecite{Koch:Theory} for a comparison of the electric current in the case of equal temperatures in the leads). For non-interacting systems the
LB result is exact. However, for the thermoelectric effects discussed in
Sec. \ref{sec:CQD} which require the presence of strong Coulomb interaction, an
alternative method such as the ME approach is needed.

We consider a single-level QD coupled to two leads $\ell\in\{A,B\}$ (such as
System 1 in Fig. \ref{fig:System} when tunnel- and Coulomb-decoupled from System
2). For simplicity, we assume energy-independent lead couplings
$\gamma^\ell(\epsilon)=\gamma^\ell$ in this case. The Hamiltonian of the QD
reduces to
\begin{equation}
\hat{H}_\text{dots}=\epsilon_1^{\phantom\dagger}\hat{c}_1^\dagger \hat{c}_1^{\phantom\dagger},
\end{equation}
with states labeled by the occupancy, $\ket{n_1}\in\{\ket{0},\ket{1}\}$.

In the LB formalism, the electric current and heat current going into lead $A$
are given by~\cite{Jauho:Kinetics,Lopez:Heat},
\begin{equation}\label{eq:LBCu}
  I_A^\text{LB}=\frac{-e}{h} \int \! d\epsilon \,
  T(\epsilon)[f^B(\epsilon) - f^A(\epsilon)],
\end{equation}
and
\begin{equation}\label{eq:LBHe}
  J_A^\text{LB} = \frac{1}{h} \int\!d\epsilon \,  
  (\epsilon - \mu_A)T(\epsilon)[f^B(\epsilon) - f^A(\epsilon)],
\end{equation}
respectively. For a non-interacting single-level QD the transmission
function $T(\epsilon)$ is
\begin{equation}
  T(\epsilon)=\frac{\gamma^A\gamma^B}{(\epsilon - \epsilon_1)^2+(\gamma/2)^2},
\end{equation}
where $\gamma= \gamma^A+\gamma^B$ and we have omitted the tunneling-induced
energy shift which is not captured by the $T$-matrix approach.

\begin{figure}[!b]
  \centering
  \includegraphics[width=\columnwidth]{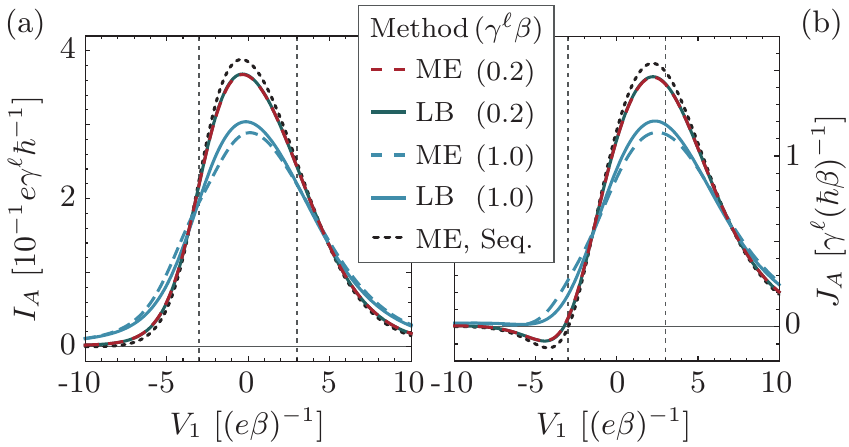}
  \caption{Comparison of the electric current (a) and heat current (b)
    calculated with the ME and LB approaches. Currents are plotted as
    function of gate voltage $V_1$ for two different lead coupling strengths
    $\gamma^{A}=\gamma^{B}=\gamma^\ell$ (energy independent). The ME result
      including only sequential tunneling is shown for reference (black
    dotted), and the vertical dashed lines mark the alignment of the dot level
    with the electrochemical potentials of lead A (left) and B
    (right). Parameters: $T_B=2T_A\equiv 2/(k_B\beta)$, $\mu_A=3\,\beta^{-1}$,
    $\mu_B=-3\,\beta^{-1}$, and $\eta=10^{-3}\,\beta^{-1}$.}
\label{fig:SingleDot_MasterLandauer}
\end{figure}

The transport currents calculated with the two approaches with a finite bias and
temperature difference ($T_B=2T_A\equiv2T$) between the leads are plotted in
Figs.~\ref{fig:SingleDot_MasterLandauer}(a)
and~\ref{fig:SingleDot_MasterLandauer}(b) as a function of the gate voltage for
two different lead coupling strengths. To demonstrate the importance of
cotunneling processes, we have included ME results based on sequential tunneling
only (black dotted curves) which do not depend on $\gamma^\ell$ in the units
shown, as well as sequential plus cotunneling (dashed curves). The results based
purely on sequential tunneling differ significantly from the LB results unless
$\gamma^\ell \ll k_B T$. However, for $\gamma^\ell < k_B T$,
the ME results with cotunneling are in excellent agreement with the LB
formalism. For $\gamma^\ell > k_B T$ which is outside the regime of validity of
the ME approach, the two approaches deviate, as expected.

In the following discussion of thermoelectric effects, the heat current is of
particular interest. As seen in Fig. \ref{fig:SingleDot_MasterLandauer}(b), when
the dot level is above the electrochemical potential in lead $A$, the heat
current becomes negative (for sufficiently small lead coupling strength). In
this case, electrons above the electrochemical potential tunnel out of the lead
and thereby cool the lead [cf. Eq.~\eqref{eq:Heat}]. Such cooling mechanisms due
to energy-selective tunneling have been confirmed experimentally in metallic QD
systems~\cite{Pekola:CoulombRefrigerator,Pekola:Onchip}. The energy-selective tunneling gives rise to an asymmetry in the energy dissipation
between the source and drain leads which was recently observed in molecular
junctions~\cite{Lee:Heat}.

\section{Thermoelectric effects in Coulomb-coupled QDs}\label{sec:CQD}

In the remaining part of the paper, we study the thermoelectric properties of
the system illustrated in Fig. \ref{fig:System}, i.e. two single-level QDs with
QD1 tunnel-coupled to leads $A$ and $B$ and QD2 tunnel-coupled to lead $C$. The
CCQD system is described by the Hamiltonian
\begin{equation}
  \label{eq:H}
  \hat{H}_\text{dots}= \epsilon_1^{\phantom\dagger}\hat{c}_1^\dagger \hat{c}_1^{\phantom\dagger} + \epsilon_2^{\phantom\dagger}\hat{c}_2^\dagger \hat{c}_2^{\phantom\dagger} + U \hat{n}_1^{\phantom\dagger} \hat{n}_2^{\phantom\dagger},
\end{equation}
where we have used the simplified notation $U_{12}\equiv U$, and the occupation states are
$\ket{m}=\ket{n_1n_2}\in\{\ket{00},\ket{10},\ket{01},\ket{11}\}$. We consider
situations where a source-drain bias $V$ is applied to System 1,
$\mu_A=\mu_0+eV/2$, $\mu_B=\mu_0-eV/2$ (we set $\mu_0=0$ as reference).

As pointed out above, we here allow for energy-dependent lead couplings. For
bias voltages and temperature differences small compared to the energy scale at
which the lead couplings vary, it suffices to consider the expansion of the lead
couplings around their value at $\mu_0$\cite{FN1},
\begin{equation}
  \label{eq:Linear}
  \gamma^\ell(\epsilon) = \gamma_0^\ell + (\epsilon-\mu_0) \partial\gamma^\ell , 
\end{equation}
where $\gamma_0^\ell=\gamma^\ell(\mu_0)$,
$\partial\gamma^\ell \equiv \tfrac{\partial\gamma^\ell(\epsilon)}
{\partial\epsilon}|_{\epsilon=\mu_0}$.

\subsection{Current and energy exchange}
\label{sec:EnergyExchange}

\begin{figure}[!t]
  \centering
  \includegraphics[width=\columnwidth]{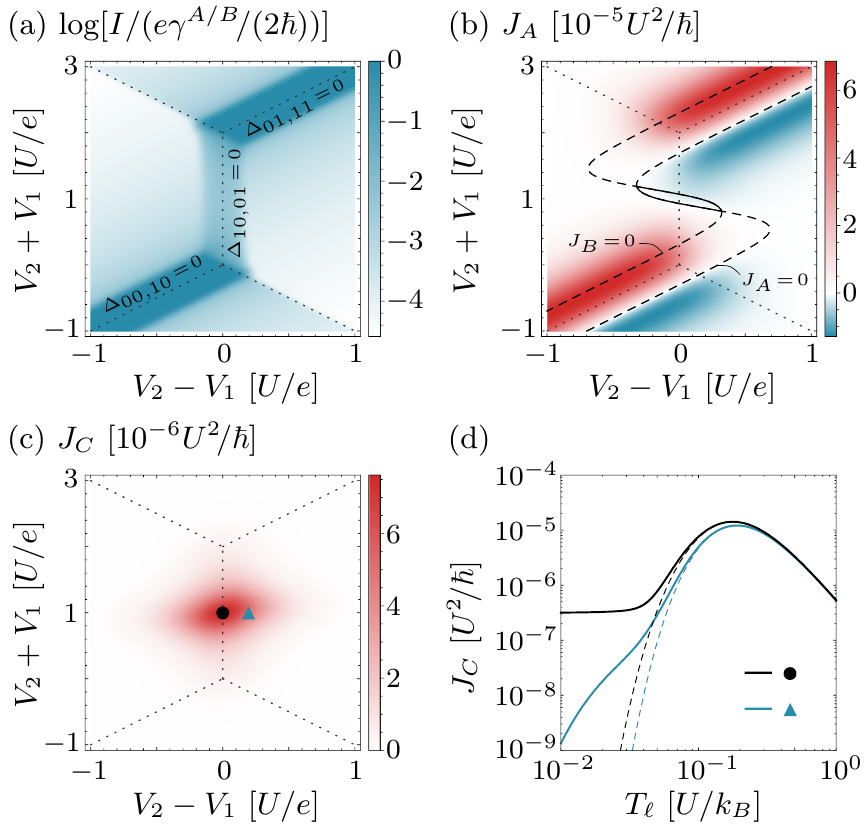}
  \caption{Electric current and heat currents. (a) Electric current in System 1 as function of gate detuning
    $V_2-V_1$ and total gating $V_2+V_1$ at low temperature,
    $k_BT_{\ell}=10^{-2}\,U$. (b) Heat current in lead $A$, $J_A$, at high
    temperature, $k_BT_\ell=10^{-1}\,U$ (contours indicate where $J_A$ and $J_B$ are
    zero). (c) Heat current in lead $C$, $J_C$, for
    $k_BT_{\ell}=10^{-1}\,U$. (d) $J_C$ as function of temperature with (solid)
    and without (dashed) cotunneling for the gate configurations marked in (c):
    $eV_{1,2}=0.5\,U$ (black circle) and $eV_1=0.4\,U$, $eV_2=0.6\,U$ (blue
    triangle). In plots (a)--(c), the degeneracy lines of the honeycomb
    vertex are indicated with dotted lines.  Parameters:
    $\gamma^{A/B}(\epsilon)=10^{-3}\,U$, $\gamma^{C}(\epsilon)=10^{-2}\,U$, and
    $eV=0.3\,U$.}
\label{fig:Cu_Heat_Cot}
\end{figure}

In Fig.~\ref{fig:Cu_Heat_Cot}(a) we show the electric current through QD1,
$I\equiv I_A=-I_B$, at low temperature $k_BT_\ell=10^{-2}U$ (for illustrative convenience) and bias $eV=0.3U$ as a function of gate detuning $V_2-V_1$ and total
gating $V_1+V_2$ in the vicinity of the honeycomb vertex of the stability
diagram~\cite{Kouwenhoven:DQD}. Here, we initially assume energy-independent
lead couplings which is sufficient to get an overall understanding of the
behavior of the system. The large current near the degeneracy lines defined by
$\Delta_{00,10}=0$ and $\Delta_{01,11}=0$ is due to sequential tunneling
processes. Away from these degeneracy lines where sequential tunneling is
exponentially suppressed, cotunneling processes give rise to a weak background
current. At the degeneracy line $\Delta_{10,01}=0$ connecting the two triple
points at $V_{1}=V_2=0,U$, respectively, \emph{nonlocal} cotunneling processes
are responsible for the enhanced cotunneling current.

The heat currents which accompany the electric current are shown in
Figs.~\ref{fig:Cu_Heat_Cot}(b)--\ref{fig:Cu_Heat_Cot}(d) for different
temperatures in the leads. Figure~\ref{fig:Cu_Heat_Cot}(b) shows the heat
current in lead $A$ for $k_BT_\ell=0.1U$. Along the degeneracy lines where
$\Delta_{00,10}=0$ and $\Delta_{01,11}=0$ and only the occupation of QD1
fluctuates, the heat current shows a behavior similar to the one in
Fig.~\ref{fig:SingleDot_MasterLandauer}(b) for a single-level QD. However, at
the center of the stability diagram, Coulomb-mediated energy exchange due to the
strong Coulomb interaction between the QDs becomes significant. This manifests
itself in a cooling of System 1 inside the region bounded by the solid lines at
the center of Fig.~\ref{fig:Cu_Heat_Cot}(b) (notice that the color scale is
dominated by the heat current with larger magnitude outside this region). From
the heat current in lead $C$ shown in Fig.~\ref{fig:Cu_Heat_Cot}(c), the cooling
of System 1 is seen to be at the cost of heating System 2. This Coulomb-mediated
energy exchange between the two QD systems occurs in spite of the fact that no
electrons are exchanged, and is the driving force behind demon-induced
cooling~\cite{Strasberg:Thermodynamics,Pekola:Onchip}, energy
harvesting~\cite{Buttiker:Optimal,Jordan:Thermoelectric,Molenkamp:Three,Molenkamp:Thermo},
and Coulomb drag~\cite{Buttiker:Mesoscopic,Jauho:Correlated}.

A simple analytical result for the energy exchange can be found when considering
sequential tunneling processes only (indicated by the superscript $s$). In this
case, the total heat currents in System 1, $J_1^s\equiv J_A^s+J_B^s$, and System
2, $J_2^s\equiv J_C^s$, become~\cite{Buttiker:Optimal}
\begin{subequations}
  \label{eq:J1JC}
  \begin{align}
    J_1^s & = \frac{U}{\tau^s} (\Gamma^+ - \Gamma^-) 
              + \frac{\mu_A-\mu_B}{e}I^s , \label{eq:J1s} \\
    J_2^s & = \frac{U}{\tau^s}(\Gamma^--\Gamma^+) \label{eq:JCs},
  \end{align}
\end{subequations}
where $\Gamma^-\equiv \Gamma_{00,01}\Gamma_{01,11}\Gamma_{11,10}\Gamma_{10,00}$,
$\Gamma^+\equiv \Gamma_{00,10}\Gamma_{10,11}\Gamma_{11,01}\Gamma_{01,00}$. The
factor $\tau^s$ depends on the various sequential tunneling rates, however, is
merely a normalization factor and is not reproduced here. The first two terms
proportional to $U$ in Eq.~\eqref{eq:J1JC} describe the energy exchange, whereas
the last term in Eq.~\eqref{eq:J1s} describes the contribution from Joule
heating in System 1. The direction of the energy transfer is determined by the
sign of $\Gamma^- - \Gamma^+$. It is therefore convenient to consider the ratio
\begin{align}
  \label{eq:GMGP}
  \frac{\Gamma^-}{\Gamma^+} = \Omega e^{U(\beta_2-\beta_1)},
\end{align}
which describes whether energy is transferred from System 1 to 2
($\Gamma^-/\Gamma^+>1$) or vice versa
($\Gamma^-/\Gamma^+<1$)~\cite{Sanchez:Detection}. On the right-hand side
of~\eqref{eq:GMGP}, we have taken $\beta_{A/B}=\beta_1$ and $\beta_C =\beta_2$,
and expressed the ratio in terms of an exponential factor, which depends on the
temperature in System 1 and System 2, and
\begin{align}
  \label{eq:DemonFac}
  \Omega \equiv \frac{(\gamma^A_1 f^A_1  +  \gamma^B_1f^B_1) 
                      (\gamma^A_0 f^A_0 e^{-\beta_1\mu_A} \! + \!
                       \gamma^B_0 f^B_0 e^{-\beta_1\mu_B})}
                     {(\gamma^A_0 f^A_0 \! + \! \gamma^B_0 f^B_0)
                      (\gamma^A_1 f^A_1 e^{-\beta_1\mu_A} \! + \!
                       \gamma^B_1 f^B_1 e^{-\beta_1\mu_B})} ,
\end{align}
which depends on the temperature and bias in System 1 only. The subscript 0 (1)
in Eq. \eqref{eq:DemonFac} indicates that the corresponding function is
evaluated at $\Delta_{00,10}$ ($\Delta_{01,11}$) [see
Eqs.~\eqref{SeqTuRates1}--\eqref{SeqTuRates2}].

The exponential factor in \eqref{eq:GMGP} shows that a temperature gradient
between the two QD systems can generate a net heat flow from the hot to the cold
system. This is the mechanism behind the heat engine studied in
Ref.~\onlinecite{Buttiker:Optimal}. On the other hand, a closer inspection of
the $\Omega$ factor reveals that it is, in fact, possible to generate a net heat
flow in the opposite direction, i.e. from the cold to the hot system, and this
is the cause of the negative heat current at the center of
Fig. \ref{fig:Cu_Heat_Cot}(b). This so-called demon-induced cooling effect will be discussed
further in Sec.~\ref{sec:Demon} below.

\begin{figure}[!t]
  \centering
  \includegraphics[width=\columnwidth]{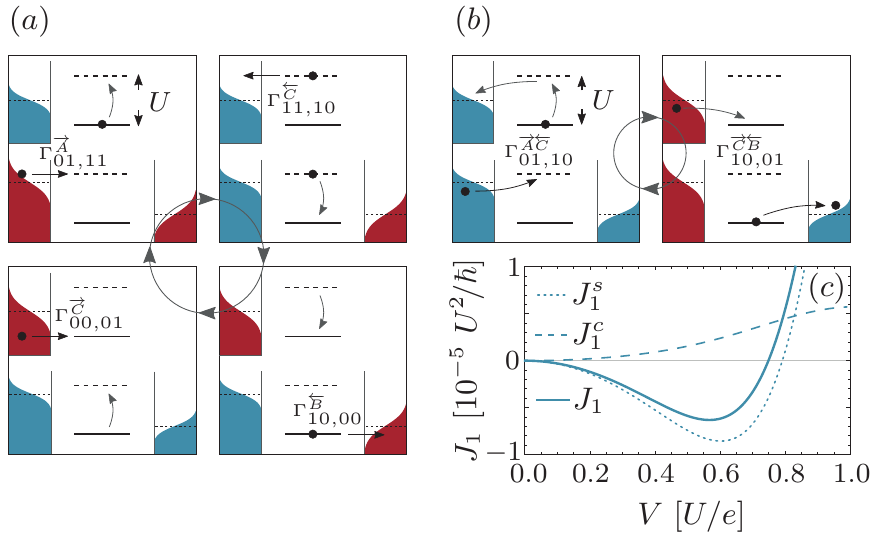}
    \caption{Cooling cycle and effect of cotunneling. (a) Sequence of sequential
    tunneling processes which cools System 1. The positions of the dot levels
    when the other dot is empty (occupied) is illustrated with solid (dotted)
    lines. (b) Sequence of nonlocal cotunneling processes. (c)
    Heat current $J_1$ as function of bias voltage. The individual contributions
    from sequential ($J_1^s$) and cotunneling ($J_1^c$) are also shown.
    Parameters: $eV_1=eV_2=U/2$, $\gamma^{A/B}(\epsilon)=10^{-3}\,U$,
    $\gamma^C(\epsilon)=10^{-2}\,U$, and $k_BT=0.1\,U$.}
\label{fig:Cycle}
\end{figure}

When the applied bias and temperature are small compared to the inter-dot
Coulomb interaction, $eV, k_B T \ll U$, cotunneling processes start to dominate
the heat currents. This is demonstrated in Fig.~\ref{fig:Cu_Heat_Cot}(d) which
shows the heat current $J_C$ as a function of temperature for the two different
gate tunings marked with symbols in Fig.~\ref{fig:Cu_Heat_Cot}(c). Considering
sequential tunneling only (dashed curves), the heat current is quenched at
$k_BT \ll U$ as $\Gamma_{01,11}$ and $\Gamma_{10,00}$ in $\Gamma^-$ become
exponentially suppressed. This can also be understood from the illustration in
Fig.~\ref{fig:Cycle}(a) which shows the sequence of sequential tunneling
processes corresponding to $\Gamma^-$. However, nonlocal cotunneling processes allow the
system to fluctuate between the two states $10\leftrightarrow 01$, as
illustrated in Fig. \ref{fig:Cycle}(b), and thereby transfer heat between the
systems. The nonlocal cotunneling channel is open for
$|\Delta_{01,10}|\lesssim \text{max}\{|eV/2|,k_BT\}$, and the associated heat
current is thus also suppressed at low temperature when $\Delta_{01,10}\neq 0$
as illustrated by the blue curve (triangle) in
Fig. \ref{fig:Cu_Heat_Cot}(d). For zero detuning $\Delta_{01,10}=0$ (circle),
the nonlocal cotunneling rates, and hence also the heat current, saturate at
$k_B T \ll eV$. In Sec. \ref{sec:Demon}, we demonstrate that nonlocal
cotunneling processes have a significant effect on the demon-induced cooling mechanism.

\subsection{Demon-induced cooling}
\label{sec:Demon}

The effect of cooling System 1 at the cost of heating System 2 has recently been
discussed in context of a Maxwell's demon where System 2 plays the role of the
demon which performs the necessary feedback to cool System
1~\cite{Strasberg:Thermodynamics,Pekola:Onchip}. To
maximize the achievable cooling power for refrigeration purposes\cite{Efficiency:Kutvonen}, large
tunneling rates, $\gamma^\ell(\epsilon)\sim k_BT, U$, are essential
[cf. Eq.~\eqref{eq:J1JC}]. However, large tunneling rates increase the
contribution from higher-order tunneling processes, thus emphasizing the
importance of including cotunneling processes in quantitative analyses of the
cooling power.

In the following, we consider the case of uniform temperature $T_\ell\equiv T$
whereby the exponential factor in \eqref{eq:GMGP} becomes unity. This allows us
to focus on the $\Omega$ factor in the optimization of the
performance. Equation~\eqref{eq:J1JC} shows that the cooling mechanism is
governed by $\Gamma^-$ since, as illustrated in Fig.~\ref{fig:Cycle}(a), in a
full sequential cycle an amount of energy $U$ is transferred from System 1 to
System 2 thereby cooling System 1. In the following, we discuss how to increase
the cooling power by maximizing the success rate for completing the cooling
cycle in Fig.~\ref{fig:Cycle}(a).

\subsubsection{Cotunneling limitations}

Although the cycle of nonlocal cotunneling processes illustrated in
Fig. \ref{fig:Cycle}(b) gives the same net transfer of electrons as the
sequential tunneling cycle in Fig. \ref{fig:Cycle}(a), the net energy transfer is different for the two cases. As illustrated, in a cotunneling process also electrons below (above)
the electrochemical potential can tunnel out of lead $A$ (into lead $B$), and
thus reduce the demon-induced cooling effect.

In Fig.~\ref{fig:Cycle}(c), we show the heat current $J_1$ together with its
individual contributions from sequential ($J_1^s$) and cotunneling ($J_1^c$)
processes. Overall, System 1 cools at low bias, while at higher bias, Joule
heating becomes dominant. The minimum in $J_1$ as a function of
bias voltage is referred to as the maximum cooling power,
$J_{1, \max}\equiv\min J_1(V)$. As the figure shows, cotunneling reduces
the maximum cooling power.
\begin{figure}[!t]
  \centering
  \includegraphics[width=\columnwidth]{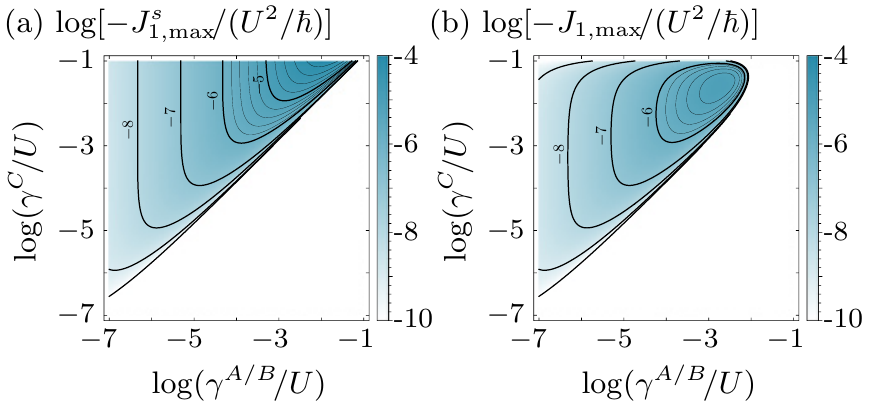}
    \caption{Maximum cooling power, $J_{1,\max}$, as function of the lead coupling
    strengths for energy-independent couplings. (a) Sequential tunneling, and
    (b) sequential plus cotunneling. Parameters: $eV_1=eV_2=U/2$ and $k_BT=0.1\,U$.}
\label{fig:GPGM}
\end{figure}

Figure~\ref{fig:GPGM} shows how the maximum cooling power $J_{1, \max}$ scales
with the lead coupling strengths. As the figure demonstrates, the rates must
satisfy $\gamma^C>\gamma^{A/B}$ to ensure that System 2 acts sufficiently fast
to perform the desired feedback such that the cooling cycle in
Fig.~\ref{fig:Cycle}(a) is completed when an electron tunnels between lead $A$
and $B$~\cite{Efficiency:Kutvonen}. In the region of large cooling power,
cotunneling processes start to become important, and hence there is a trade-off
between sequential tunneling which improves the cooling effect, and nonlocal
cotunneling which limits the effect. In addition, the area in the lead coupling
parameter space where refrigeration is possible is also reduced when cotunneling
is included.

\subsubsection{Performance boosting}

Here we demonstrate that energy-dependent lead couplings
can enhance the demon-induced cooling power significantly. We restrict the
discussion to lead couplings with a linear energy dependence
[cf. Eq.~\eqref{eq:Linear}].

\begin{figure}[!b]
  \centering
  \includegraphics[width=0.75\columnwidth]{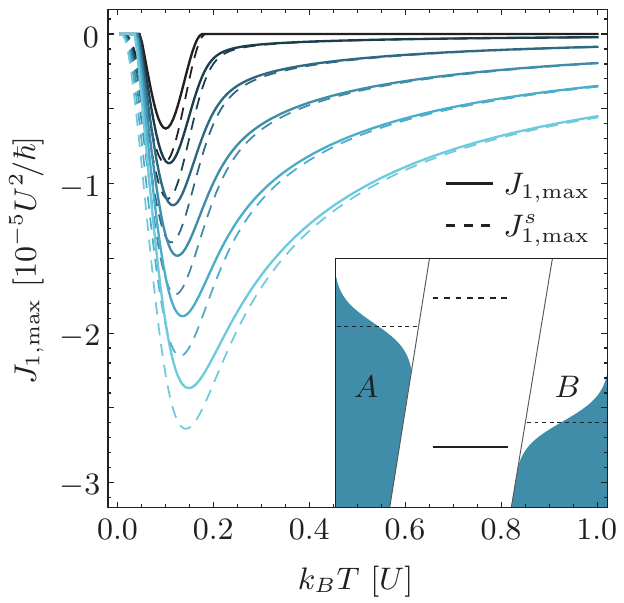}
    \caption{Performance boosting with energy-dependent lead couplings. Maximum
    cooling power as function of temperature for different lead coupling
    strengths: $\partial\gamma^A=-\partial\gamma^B=x \gamma_0^{A/B}/U$ (sketched
    in the inset), with $x=0$ (black) to $x=1$ (light blue) in steps of
    $0.2$. The full (dashed) lines show the result obtained with (without)
    cotunneling. Parameters: $\gamma^C(\epsilon)=10^{-2}\,U$,
    $\gamma_0^{A/B}=10^{-3}\,U$, $eV_1=eV_2=U/2$, and $\eta=10^{-4}\,U$.}
\label{fig:6}
\end{figure}

By inspecting the $\Omega$ factor in Eq.~\eqref{eq:DemonFac}, we find that for
$\mu_A>\mu_B$, the configuration illustrated in the inset of Fig.~\ref{fig:6}
where $\gamma^A_0$, $\gamma^B_1$ are reduced compared to $\gamma^A_1$,
$\gamma^B_0$, boosts the $\Omega$ factor (and thereby
$\Gamma^-/\Gamma^+$). This results in an enhancement of the cooling power
by suppressing direct tunneling between lead $A$ and $B$ via two sequential
tunneling processes which contributes to Joule heating, while at the same time
promoting the processes of the cooling cycle in Fig.~\ref{fig:Cycle}(a).

In Fig.~\ref{fig:6} we show the maximum cooling power as a function of
temperature for different situations for the energy dependence of the lead
couplings, from the top (black) curve showing the result for energy \emph{independent} lead couplings, to increasing energy dependence,
i.e. increasing $\abs{\partial\gamma^{A/B}}$, towards the bottom (light blue)
curve. When tuning the energy dependence of the lead couplings, a significant
enhancement of the cooling power is achieved. Again, the effect of cotunneling
processes is to reduce the attainable cooling power (solid lines) relative to
the cooling power obtained when only considering sequential tunneling processes
(dashed lines).

\section{Conclusions}\label{sec:conc}

In summary, we have studied thermoelectric effects in CCQD systems with a $T$-matrix based master-equation approach
for the calculation of charge and heat currents. Importantly, our method (i)
treats \emph{incoherent} sequential tunneling processes and \emph{coherent}
cotunneling processes on equal footing, and (ii) can account for energy-dependent
tunnel couplings to the leads. Both are essential for quantitative predictions
and optimization of the thermoelectric properties of CCQDs.

To benchmark the master-equation method, we considered a non-interacting
single-level QD coupled to source and drain leads for which the
Landauer-B{\"u}ttiker formalism is exact. In the regime of validity of our
method, i.e. small tunnel couplings to the leads, $\gamma< k_BT$, we
demonstrated excellent agreement with the results from the Landauer-B{\"u}ttiker
method when cotunneling processes are included in the master equation.

Furthermore, we studied the effect of cotunneling processes and energy-dependent
lead couplings on the thermoelectric properties of a CCQD system consisting of
two QDs exhibiting a Maxwell's demon-like cooling
mechanism~\cite{Strasberg:Thermodynamics,Pekola:Onchip}. First of all, we
showed that cotunneling processes reduce the cooling effect since
cotunneling processes do not share the delicate energy selectivity inherent
  to sequential tunneling processes. This results in a significant reduction of
the achievable cooling power compared to the sequential tunneling result when the lead couplings are increased to
  maximize the cooling power from sequential tunneling processes. Secondly, we
demonstrated that it is possible to boost the cooling power significantly via
other means by introducing energy-dependent lead couplings and properly tuning
their energy dependence. In this case, we showed that cotunneling still
reduces the cooling power significantly, thus emphasizing the importance of
cotunneling processes in quantitative analyses. 

Applying the methodology to
other mesoscopic systems allows for testing of new thermoelectric device ideas
beyond sequential tunneling estimates, as well as for improved comparison with
experiments.

\begin{acknowledgments}
  We would like to thank J.~P. Pekola, M. Leijnse, C. Timm, and N.~M. Gergs for valuable
  discussions. K.K. acknowledges support from the European Union's Horizon
  2020 research and innovation programme under the Marie Sklodowska-Curie grant
  agreement no.~713683 (COFUNDfellowsDTU). The Center for Nanostructured
Graphene (CNG) is sponsored by the Danish Research Foundation, Project DNRF103.
\end{acknowledgments}

\appendix

\section{Cotunneling rates and regularization procedure} 
\label{sec:regularization}

The rate for elastic cotunneling through a single-level QD is given by
\begin{align}
  \label{eq:Cotunneling1elas}
  \tilde{\Gamma}^{\small{\overrightarrow{\ell\phantom{'}}\overleftarrow{\ell'}}}_{mm}
  & = \int\! \frac{d\epsilon}{2\pi\hbar}
      \gamma^{\ell}(\epsilon) \gamma^{\ell'}\!(\epsilon)
      f^{\ell\phantom{'}}\!(\epsilon)\bar{f}^{\ell'}\!(\epsilon)
    \left|\frac{1}{\Delta_{vm} \pm \epsilon + i\eta}\right|^2\!\!\!,
\end{align}
where $v$ refers to the virtually occupied intermediate state created in the
process where an initially empty level is filled ($+\epsilon$) or an initially
filled level is emptied ($-\epsilon$).

In pair-cotunneling processes, two electrons tunnel simultaneously out of (into)
the QD system and into (out of) the leads $\ell$ and $\ell'$. The rate for such
processes takes the form
\begin{align}
\label{eq:Cotunneling2}
\tilde{\Gamma}^{\small{\overleftarrow{\ell\phantom{'}}\overleftarrow{\ell'}}}_{mn}
  &= \int\! \frac{d\epsilon}{2\pi\hbar}     
     \gamma^\ell(\epsilon) \gamma^{\ell'}\!(\Delta_{nm} - \epsilon)
    \bar{f}^\ell(\epsilon) \bar{f}^{\ell'}\!(\Delta_{nm} - \epsilon)
     \nonumber\\
  &\quad\times \left|\frac{1}{\Delta_{vm} - \epsilon + i\eta} +
                     \frac{1}{\Delta_{v'n} + \epsilon + i\eta}\right|^2\!\!\!,
\end{align}
where $v$ ($v'$) refers to the virtually occupied intermediate state in a
process where an electron initially tunnels from the QD system and into lead
$\ell$ ($\ell'$). Similarly,
\begin{align}
\label{eq:Cotunneling3}
\tilde{\Gamma}^{\small{\overrightarrow{\ell\phantom{'}}\overrightarrow{\ell'}}}_{mn}\!&=\!\!\int\!\!\frac{d\epsilon}{2\pi\hbar}\gamma^\ell(\epsilon)\gamma^{\ell'}\!(\Delta_{mn} - \epsilon)f^\ell(\epsilon)f^{\ell'}\!(\Delta_{mn} - \epsilon)\nonumber\\
&\quad\times\!\left|\frac{1}{\Delta_{vn} - \epsilon + i\eta}+\frac{1}{\Delta_{v'm}\!+\epsilon + i\eta}\right|^2\!\!\!,
\end{align}
where $v$ ($v'$) refer to the virtually occupied intermediate state in a process
where an electron initially tunnels from lead $\ell'$ ($\ell$) and into the
QD system.

The bare cotunneling rates are formally divergent in the limit $\eta\to0$. The divergence stems from factors involving
$|x+i\eta|^{-2}$, $x,\eta\in\mathbb{R}$. Using that~\cite{Matveev:Cotunneling}
\begin{align}
  \label{eq:XXX}
  \left\vert \frac{1}{x + i \eta} \right\vert^2 
  \to \frac{\pi}{\eta} \delta(x) + \mathcal{P}\frac{1}{x^2}  , \quad \eta\to 0^+,
\end{align}
where $\mathcal{P}$ denotes the principle value, we can identify the divergent contributions, e.g. from Eq. \eqref{eq:Cotunneling1}
\begin{equation}
  \label{eq:RegRates}
  \tilde{\Gamma}_{mn}^{\small{\overrightarrow{\ell\phantom{'}}\overleftarrow{\ell'}}}\to\frac{\hbar}{2\eta}\left(\Gamma^{\small{\overrightarrow{\ell\phantom{'}}}}_{mv}\Gamma^{\small{\overleftarrow{\ell'}}}_{vn}+\Gamma^{\small{\overleftarrow{\ell'}}}_{mv'}\Gamma^{\small{\overrightarrow{\ell\phantom{'}}}}_{v'n}\right)+\Gamma_{mn}^{\small{\overrightarrow{\ell\phantom{'}}\overleftarrow{\ell'}}},
\end{equation}
where
$\Gamma_{mn}^{\small{\overrightarrow{\ell\phantom{'}}\overleftarrow{\ell'}}}$
denotes the regularized cotunneling rate, and we have used that the cross-terms
from the absolute squared in Eq. \eqref{eq:Cotunneling1} do not contribute to any
divergences. The divergent contribution
is proportional to products of two sequential tunneling rates. These correspond
to two energy-conserving (sequential) transitions which can be identified with
the intermediate processes in the cotunneling process. The sum is over the
possible sequences of intermediate transitions. Similarly, for the cotunneling
heat rates, e.g. Eq. \eqref{eq:CotunnelingHeat1}
\begin{equation}
  \label{eq:RegHeat}
  \tilde{W}_{\ell,mn}^{\small{\overrightarrow{\ell\phantom{'}}\overleftarrow{\ell'}}}\!\!\to\!\frac{\hbar}{2\eta}\!\left[W^{\small{\overrightarrow{\ell\phantom{'}}}}_{\ell, mv}\Gamma^{\small{\overleftarrow{\ell'}}}_{vn} + \Gamma^{\small{\overleftarrow{\ell'}}}_{mv'}W^{\small{\overrightarrow{\ell\phantom{'}}}}_{\ell,v'n}\right]+W_{\ell,mn}^{\small{\overrightarrow{\ell\phantom{'}}\overleftarrow{\ell'}}},
\end{equation}
or the corresponding heat rate in lead $\ell'$
\begin{equation}
  \label{eq:RegHeat2}
  \tilde{W}_{\ell',mn}^{\small{\overrightarrow{\ell\phantom{'}}\overleftarrow{\ell'}}}\!\!\to\!\frac{\hbar}{2\eta}\!\left[\Gamma^{\small{\overrightarrow{\ell\phantom{'}}}}_{mv}W^{\small{\overleftarrow{\ell'}}}_{\ell',vn} + W^{\small{\overleftarrow{\ell'}}}_{\ell', mv'}\Gamma^{\small{\overrightarrow{\ell\phantom{'}}}}_{v'n}\right]+W_{\ell',mn}^{\small{\overrightarrow{\ell\phantom{'}}\overleftarrow{\ell'}}}.
\end{equation}
We apply the regularization scheme in Ref.~\onlinecite{Matveev:Cotunneling} and subtract these terms scaling as $\eta^{-1}$.

In the case of identical temperatures in the leads, using the identity
$f(\epsilon_1)[1-f(\epsilon_2)]=n(\epsilon_1-\epsilon_2)[f(\epsilon_2)-f(\epsilon_1)]$,
where $f(\epsilon)$ is the Fermi-Dirac distribution and $n(\epsilon)$ is the
Bose-Einstein distribution, the cotunneling rates can be written in the form
\begin{align}
  I & = \int_{-\infty}^{\infty}d\epsilon P(\epsilon)  
      \left[ f^{\ell'}(\epsilon) - f^\ell(\epsilon+\Delta_3) \right] \nonumber \\
    & \quad \times\left| \frac{{k_1}}{\epsilon-\Delta_{1}+i\eta} 
          + \frac{{k_2}}{\Delta_{2}-\epsilon+i\eta}\right|^2,
\end{align}
where $P(\epsilon)$ is assumed to be a polynomial,
$P(\epsilon)=\sum_{i=0}^nc_i\epsilon^n$, of maximum order $n=2$ for
$k_1-k_2\neq 0$ and $n=4$ for $k_1-k_2=0$ to ensure that the result below is
well-defined. The derivation is in line with the one in
Ref.~\onlinecite{Jauho:Correlated}, and the integral becomes \small
\begin{widetext}
\begin{equation}\label{FinalInt}
\begin{split}
I=&k_1^2P'(\Delta_1)\text{Re}\left[\psi_{\ell'}^-(\Delta_1)-\psi_{\ell}^-(\Delta_1+\Delta_3)\right]+\frac{k_1^2\beta}{2\pi}P(\Delta_1)\text{Im}\left[\psi_{1_{\ell'}}^-(\Delta_1)-\psi_{1_{\ell}}^-(\Delta_1+\Delta_3)\right]\\
&+k_2^2P'(\Delta_2)\text{Re}\left[\psi_{\ell'}^-(\Delta_2)-\psi_{\ell}^-(\Delta_2+\Delta_3)\right]+\frac{k_2^2\beta}{2\pi}P(\Delta_2)\text{Im}\left[\psi_{1_{\ell'}}^-(\Delta_2)-\psi_{1_{\ell}}^-(\Delta_2+\Delta_3)\right]\\
&-\frac{2{k_1}{k_2}}{\Delta_1-\Delta_2}\left(P(\Delta_1)\text{Re}\left[\psi_{\ell'}^-(\Delta_1)-\psi_{\ell}^-(\Delta_1+\Delta_3)\right]-P(\Delta_2)\text{Re}\left[\psi_{\ell'}^-(\Delta_2)-\psi_{\ell}^-(\Delta_2+\Delta_3)\right]\right)+R+\mathcal{O}(\eta^{-1})+\mathcal{O}(\eta),
\end{split}
\end{equation}
\end{widetext}
\normalsize
where
\begin{equation}
{\psi_{(1)}}_{\ell}^\pm(\epsilon)\equiv\psi_{(1)}\!\left(\frac{1}{2}\pm i\frac{\beta}{2\pi}(\epsilon-\mu_{\ell})\right),
\end{equation}
with $\psi$ ($\psi_1$) being the digamma (trigamma) function, and
\small
\begin{equation}
R=\left\{
\begin{array}{l}
c_2(\mu_{\ell'}-\mu_{\ell}+\Delta_3)({k_1}-{k_2})^2,\ {k_1}-{k_2}\neq0,\\
c_4(\mu_{\ell'}-\mu_{\ell}+\Delta_3)k_1^2(\Delta_1-\Delta_2)^2,\ {k_1}-{k_2}=0.\!\!
\end{array}
\right.
\end{equation}
\normalsize The term $\mathcal{O}(\eta^{-1})$ is omitted by regularization
before letting $\eta\to0$. For $k_BT<\gamma$ (outside the regime of validity), the failure of the approach is seen as a logarithmic divergence of the digamma functions near the degeneracy points.

In studies of thermoelectric effects where different lead temperatures as well
as more general energy-dependence of the lead couplings become relevant, one
must turn to a numerical procedure. In this case, we evaluate the cotunneling integrals numerically with a small
but finite $\eta$, and subsequently subtract contributions of order $\eta^{-1}$
as shown in e.g. Eqs.~\eqref{eq:RegRates}--\eqref{eq:RegHeat2}. In particular, we have applied the numerical procedure in Fig. \ref{fig:SingleDot_MasterLandauer} and Fig. \ref{fig:6}, and stated the values of $\eta$ in the figure caption.

\bibliography{journalabbreviations,references}

\begin{thebibliography}{49}
\bibitem [{FN2()}]{FN2}%
  \BibitemOpen
  \href@noop {} {}\bibinfo {note} {Notice that for the particular
  pair-cotunneling processes with $\ell=\ell'$ one should include a factor of
  two in the current, however, for single-level QDs discussed here such
  processes do not contribute.}\BibitemShut {Stop}%
\bibitem [{FN1()}]{FN1}%
  \BibitemOpen
  \href@noop {} {}\bibinfo {note} {To ensure that the lead coupling strengths
  are positive, a linear expansion is only appropriate when the bias window or
  the thermal window exponentially suppress the contribution to the cotunneling
  integrals at energies where the lead coupling becomes negative. In the
  numerical calculation we take the absolute value of the lead
  couplings.}\BibitemShut {Stop}%
\end{thebibliography}


\end{document}